# First ex-vivo positronium imaging of tissues with modular J-PET scanner using $^{44}Sc$ radionuclide.

Karol Kubat[1,2*], Manish Das[1,2], Sushil Sharma[1,2], Ermias Y. Beyene[1,2], Aleksander Bilewicz[4], Jarosław Choiński[5], Neha Chug[1,2] , Catalina Curceanu[7], Eryk Czerwiński[1,2], Jakub Hajduga[3], Sharareh Jalali[1,2], Krzysztof Kacprzak[1,2], Tevfik Kaplanoglu[1,2], Łukasz Kapłon[1,2], Kamila Kasperska [1,2], Aleksander Khreptak[1,2], Grzegorz Korcyl[1,2], Tomasz Kozik [1,2], Anoop Kunimmal Venadan[1,2] , Deepak Kumar[1,2], Sumit Kumar Kundu[1,2], Bartosz Leszczyński[1,2], Edward Lisowski[6], Filip Lisowski[6], Justyna Mędrala-Sowa[1,2], Simbarashe Moyo[1,2], Wiktor Mryka[1,2], Szymon Niedźwiecki[1,2], Anand Pandey[1,2], Piyush Pandey[1,2], Szymon Parzych[1,2], Alessio Porcelli[1,2], Bartłomiej Rachwał[3], Martin Rädler[1,2], Magdalena Skurzok[1,2], Anna Stolarz[5], Tomasz Szumlak[3], Satyam Tiwari[1,2], Pooja Tanty[1,2], Keyvan Tayefi Ardebili[1,2], Kavya Valsan Eliyan[1,2], Rafał Walczak[4], Pawel Moskal[1,2], Ewa Ł. Stępień[1,2*]

[1]Faculty of Physics, Astronomy, and Applied Computer Science, Jagiellonian University, S. Łojasiewicza 11, 30-348 Kraków, Poland

[2] Center for Theranostics, Jagiellonian University, Kopernika 40, 31-501 Kraków, Poland

[3]AGH University of Science and Technology, Poland

[4]Institute of Nuclear Chemistry and Technology, Warsaw, Poland

[5]Heavy Ion Laboratory, University of Warsaw, Warsaw, Poland

[6]Cracow University of Technology, 31-864 Kraków, Poland.

[7]INFN, Laboratori Nazionali di Frascati CP 13, Via E. Fermi 40, 00044, Frascati, Italy.

*Corresponding Authors: E-mail: karol.kubat@uj.edu.pl; e.stepien@uj.edu.pl

**Abstract**

This study presents the first ex-vivo positronium imaging of human tissues using the modular J-PET scanner with the $^{44}$Sc radionuclide. The $^{44}$Sc isotope was produced via the $^{44}$Ca(p, n)$^{44}$Sc nuclear reaction and used to perform positronium imaging of phantom composed of human adipose tissue, cardiac myxoma tissue, thrombi blood clot, and also porous polymer XAD4, and a certified reference material (CRM) made from fused silica. The experiment demonstrates the suitability of $^{44}$Sc as a positron source for positronium imaging. The performance of J-PET for positronium imaging with $^{44}$Sc was validated by proper reconstruction of the mean ortho-positronium lifetime for CRM material and XAD-4 polymer. The mean ortho-positronium (o-Ps) lifetimes determined for adipose tissue, cardiac myxoma tissues and thrombi were consistent with results of previous experiments. The study highlights the potential $^{44}$Sc radionuclide for positronium lifetime imaging (PLI).

## 1. Introduction

The PET imaging procedure begins with the intravenous administration of a radiopharmaceutical to a patient. This pharmaceutical is meticulously combined with a positron-emitting isotope, such as $^{18}$F-labelled fluoro-2-deoxyglucose ($^{18}$F-FDG), Ga[$^{68}$Ga]-DOTA-TATE, or Ga[$^{68}$Ga]-PSMA. This complex then selectively spreads throughout various tissues, transported by its carrier molecule, and acts on specific biological targets indicative of disease states [1].

The positron emitted from the nucleus travels through matter, gradually losing kinetic energy through collisions. Its range depends on the isotope and material and varies from 0.56 to 3.6 mm [2]. Eventually, the positron encounters and annihilates an electron present in the tissue-forming macromolecules. This annihilation reaction is a fundamental physical phenomenon that results in the simultaneous production of two 511-KeV photons. Importantly, these two photons are emitted at an angle of close to 180° to each other, a direct consequence of the conservation of momentum and boost of the annihilating system [1]. If annihilation does not occur instantly, the temporal state called Positronium is formed. Positronium is an exotic, hydrogen-like atom consisting of an electron (e-) and its antiparticle, a positron (e+), bound together. Unlike normal atoms, it does not have a nucleus and both particles orbit a common center of mass. It is unstable and eventually annihilates into gamma-ray photons. Depending on spin, it exists in two distinct ground states [3, 4]:

Para positronium (p-Ps)

Singlet state (1S0) with a total spin of S=0. It decays primarily into two gamma photons due to the conservation of charge conjugation symmetry. It is very short-lived, with a mean lifetime of approximately 125 picoseconds. It occurs in about 25% of cases.

Ortho positronium (o-Ps)

Triplet state (3S1) with a total spin of S=1. It decays primarily into 3 gamma photons, decay into 2 photons is forbidden by charge conjugation symmetry. It has a much longer lifetime, with a mean lifetime of about 142 nanoseconds. It arises in about 75% of cases.

During a standard PET scan, a significant proportion, as much as 40% of positron annihilation events within the patient's body does not occur through direct positron-electron collision but rather through the initial formation of positronium atoms [3, 4]. Conventional PET scanners are primarily designed to detect two 511 keV photons, which are produced by direct positron-electron annihilation, the self-annihilation of para-positronium, or the "pickoff" annihilation of ortho-positronium. The additional information contained in the dynamics of positronium formation and, most importantly, its specific lifetime within the tissue, has historically remained unexploited in routine clinical PET diagnostics [3].

A *positronium imaging* method has recently been invented [5]. This method enables us to depict properties of positronium atoms in the living organism. In particular the mean ortho-positronium lifetime and the ratio of 3 gamma to 2 gamma decay rates of ortho-positronium are considered as promising diagnostic indicators since their changes depend on the tissue molecular arrangement and oxygen concentration [5–9]. Positronium lifetime images obtained *in-vitro* [10], first *ex-vivo* [11, 12] and *in-vivo* [13] using a multi-photon J-PET scanner demonstrated that mean ortho-positronium lifetime, and also mean positron lifetime differ between healthy and tumor tissues. This observation opens perspective for application of positronium imaging in clinics [9].

The development of specialized PET scanners and the adoption of commercial systems are crucial for advancing positronium lifetime measurements, particularly in clinical application The Jagiellonian-PET (J-PET) scanner is a notable example, designed with plastic scintillators [14–16] and multi-photon detection capabilities [12, 13, 17, 18]. It played a key role in demonstrating the first *ex-vivo* and *in-vivo* positronium lifetime images in humans, highlighting its ability to simultaneously register annihilation photons and prompt gamma rays. J-PET technology is also considered potentially cost-effective, offering a significant reduction (approximately five-fold) in cost compared to crystal-based systems for building total-body PET scanners [4].

Positronium lifetime has the potential to serve as a multi-faceted diagnostic marker, offering unique and complementary information beyond that provided by traditional PET. While conventional PET excellently mapping the metabolic activity and radiotracer distribution, providing functional assessment, positronium lifetime imaging explores the submolecular architecture and local chemical environment of tissues [9]. The sensitivity of o-Ps lifetime to free volume and oxygen concentration means it can detect subtle changes in tissue microstructure that may precede or accompany metabolic alterations. The first *in-vivo* positronium images recently presented by the J-PET collaboration [19] encourages further development towards transferring this method to the clinic. Next steps require upgrading commercial scanners to multi-photon imaging capabilities [20–22], using radionuclides with a high $\beta^{+}\,\gamma$ branching ratio [1, 23–25], and developing iterative methods for positronium imaging reconstruction [26–30].

Recently, several groups have conducted positronium studies to assess their feasibility for exploring the clinical use of positronium [7, 12, 13, 20–22, 27–29, 31–38] and various advanced reconstruction techniques has also been reported [27, 29, 32, 36–38]. So far isotopes such as $^{22}$Na[12], $^{68}$Ga[13], $^{124}$I[21] and, $^{82}$Rb[22] have been explored for the positronium imaging. The long half-life of $^{22}$Na, 2.6 years, makes it unsuitable for clinical positronium lifetime imaging studies, while the low prompt photon branching ratio of $^{68}$Ga, $^{124}$I and $^{82}$Rb makes positronium imaging difficult [23, 39].

In contrast, $^{44}$Sc appears to be the best candidate for positronium imaging [9, 23, 24] with its clinically favorable half-life of 4 hours, 2 minutes and 24 seconds [40] and, most importantly it undergoes $\beta^{+}$ decay in approximately 94.3% and subsequently emits a 1157 keV prompt $\gamma$-ray with a yield of around 100%. Recently, $^{44}$Sc was used to prove the first positronium lifetime studies with phantoms [29, 41].

The radionuclide $^{44}$Sc is currently not used in clinical practice, but numerous phantom studies, preclinical studies, and *in-vivo* imaging with human shows it potential for clinical use [40, 42–59].

In this work we focus on ex-vivo analysis of tissues using positronium annihilation lifetime spectroscopy with the radioisotope $^{44}$Sc, based on measurement data obtained from the J-PET scanner. These studies were performed on phantoms consisting of biological, biochemical, and material samples. These experiments were conducted to demonstrate suitability of the radionuclide $^{44}$Sc as a positron source for positronium imaging.

## 2. Materials and Methods

### 2.1. Isotope preparation.

Nuclear reaction $^{44}$Ca (p, n) $^{44}$Sc was used for production of $^{44}$Sc isotope in Heavy Ion Laboratory at the University of Warsaw (HIL). For the production of $^{44}$Sc (as described in [60] , a calcium carbonate tablet was inserted into the graphite bed (to enhance the dissipation of the heat deposited by the p beam). The target was bombarded for 2 hours and 25 minutes with average current 10.6 mA. Preparation of a solution containing scandium radionuclide from the target material was performed in three steps under fume hood lined with liquid-absorbing mats, using lead tube containers and lead bricks placed around for radiation protection:

1. Dissolving the target disc in diluted hydrochloric acid (HCl) to obtain a scandium solution.
2. Filtering the scandium solution to remove any remaining black graphite dust.
3. Buffering the scandium solution to increase pH from 0.5 to about 4.0.

Both calcium carbonate (the target matrix) and resulting scandium carbonate are converted to water soluble chlorides when react with HCl, while carbonate group is released as $CO_2$. Graphite does not dissolve in acid solution, but small fragments may escape into the solution. Filtering is necessary to remove black graphite dust sticking to all surfaces. Raising the pH (buffering) is necessary to prevent corrosion of the steel needles used to draw up the solution and damage to the plastic phantoms. However, the pH must remain acidic (below 7) to prevent scandium from precipitating from the solution.

2.2. Phantoms composition and preparation.

This study was focused on the identification of o-Ps lifetime imaging in biological samples using $^{44}$Sc. $^{44}$Sc with an activity of 7.085 MBq was transferred in 0.25 mL aliquots into 5 samples: human adipose tissue, human cardiac myxoma tissue, human thrombi, Amberlite XAD4 and CRM No_40 made of fused silica (Certified Reference Material with known o-Ps lifetime). XAD4 was used as an inorganic sample with a long o-Ps lifetime and the CRM with a known o-Ps lifetime was used to validate the calibration of setup. The tissue and XAD4 samples were placed into 5 ml Eppendorf tubes and $^{44}$Sc solution was placed between samples in PCR tube (0.2 ml) shown in Figure 1. Human tissues were obtained from the previous study approved by the Bioethical Commission of the Jagiellonian University (No. 1072.6120.123.2017) and Ethics Committee of Krakowski Specialist Hospital named after St. John Paul II (Protocol #2023-152).

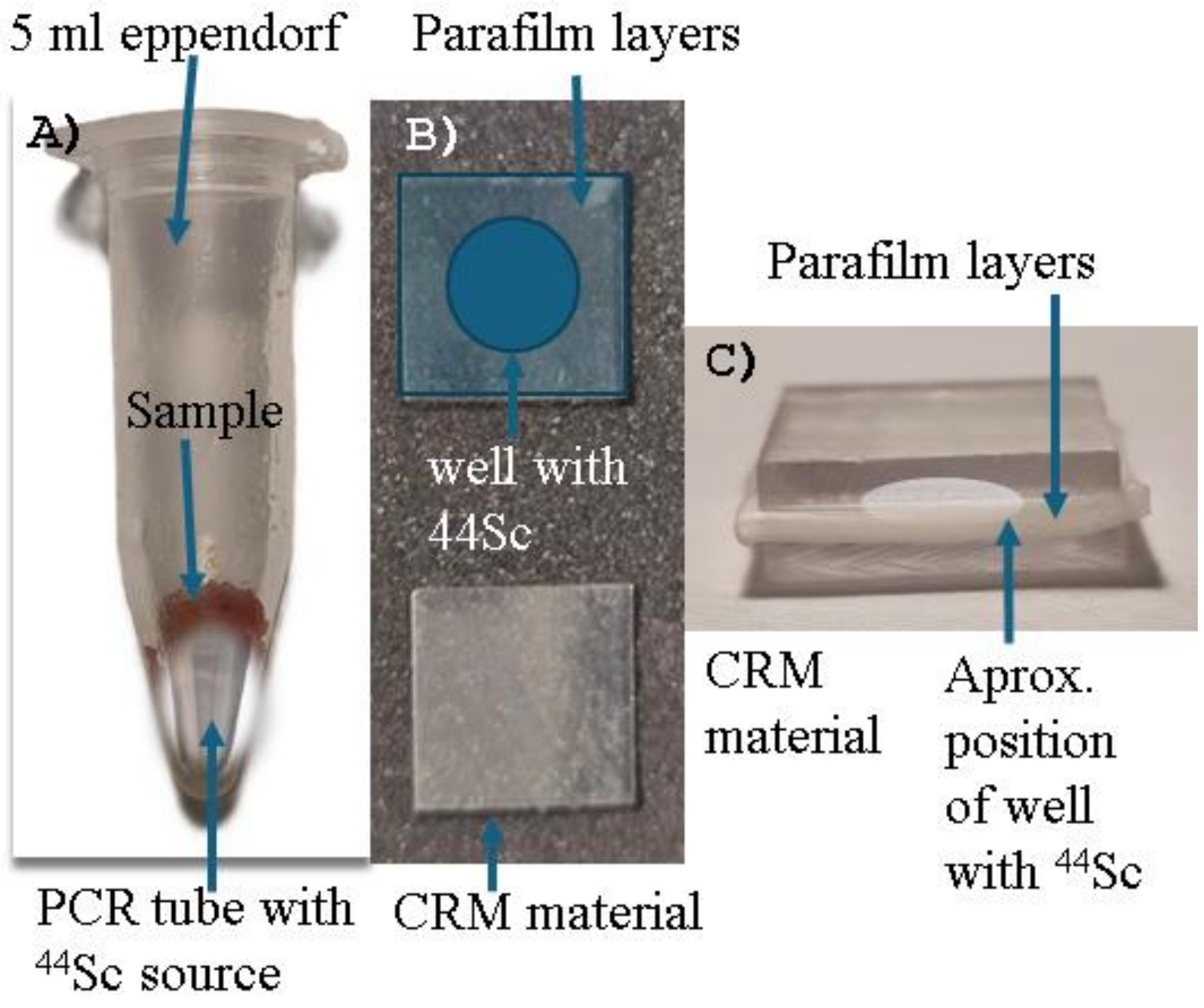

Figure 1. Photo showing the placement of the $^{44}$Sc source in phantom samples.

Placement of source inside a biological phantom (A) and between CRM material in a well cut inside layers of parafilm from top (B) and side (C).

CRM with dimensions (15 × 15 × 1.5) $mm^3$ and weight of 0.743(1) g were placed at top and bottom of source which was pipetted at bottom part into 1 cm in diameter and 0.8 cm in height well made from parafilm. Whole CRM setup was covered with parafilm layer to prevent any leakage. CRMs were obtained from National Institute of Advanced Industrial Science and Technology (AIST) in Tokyo, Japan [61]. Sample containers were placed inside the J-PET detector as shown in Figure 2.

2.3 Measurements using a modular J-PET tomograph.

The Modular J-PET represents the newest advancement in the Jagiellonian-PET series, utilizing extended plastic scintillator strips. The modular design of this prototype enables cost-effective multi-photon annihilation and positronium imaging, allowing for easy assembly, portability, and versatility. The entire detector weights approximately 60 kg. Furthermore, its lightweight design facilitates examinations on a static bed with a mobile detection system that can be conveniently positioned next to the patient, eliminating the need for spacious examination rooms. Consisting of 24 modules arranged in regular 24-sided polygons in a circle with a diameter of 73.9 cm, each module consists of 13 scintillation strips measuring 50 cm in length and 6 mm × 24 mm in cross-section[62]. The detector with positioned samples is presented in Figure 2. Measurements were performed for 6 hours, allowing for the collection of approximately 3.97 mln events. The coordinates of phantom placement are presented in Table 1. The measurement was performed in two runs: first part with CRM and second part of without CRM, each lasting 3 hours.

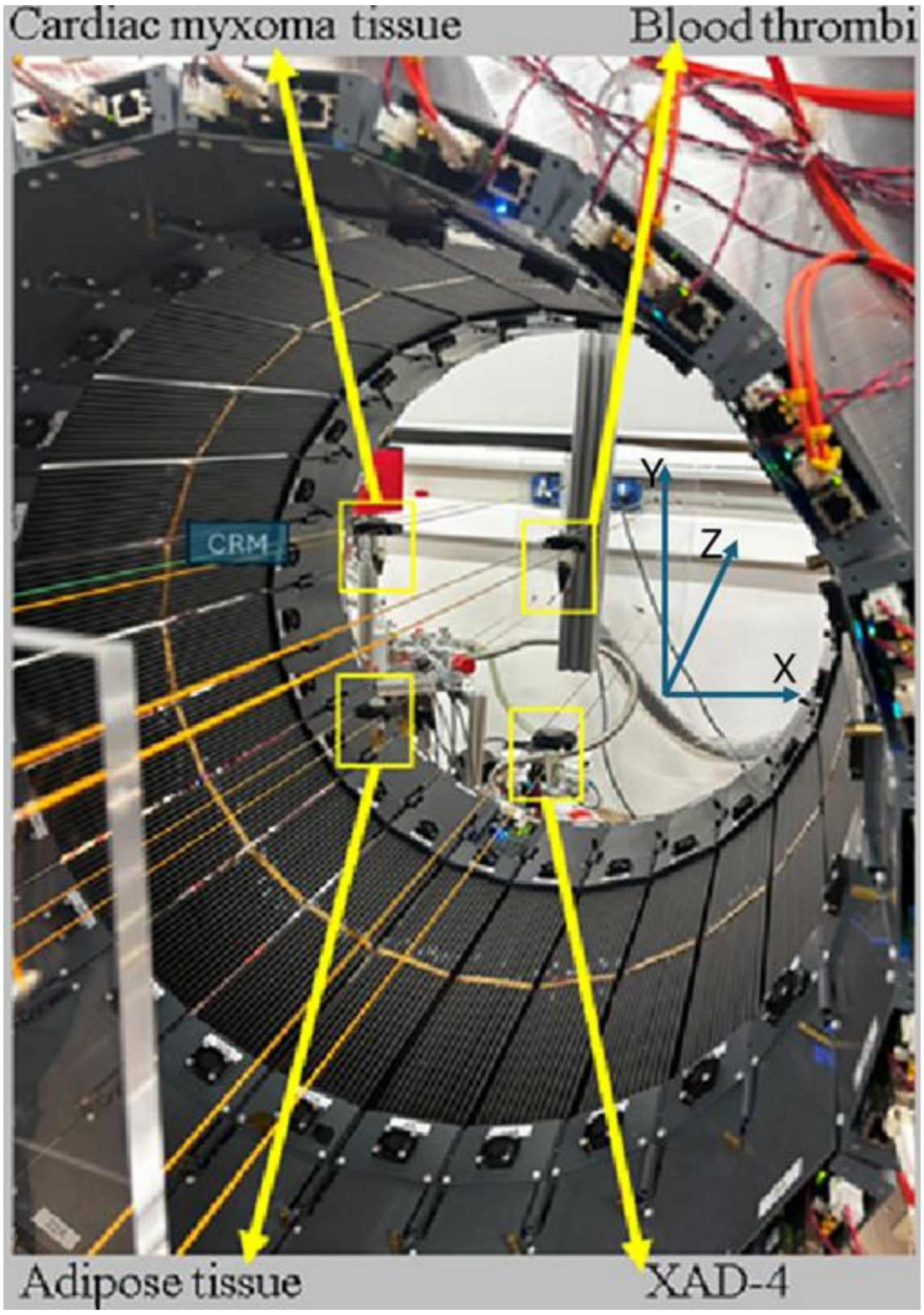

Figure 2. Photo showing the placement of phantom samples in the J-PET system.

The yellow rectangles indicate their position in the scanner.

Table 1. Placement position of samples in detector.

| Phantom | X [cm] | Y [cm] | Z [cm] |
|---|---|---|---|
| Adipose tissue | -10 | -10 | 0 |
| Cardiac myxoma tissue | -10 | 10 | 0 |
| Thrombin blood clot | 10 | 10 | 0 |
| Dry XAD 4 | 10 | -10 | 0 |
| CRM | -10 | 14 | 15 |

2.4. Micro- CT reconstruction.

The biological phantoms used in the experiment, were scanned to show position of PCR tube. Prior to micro-CT scanning, the samples were stained with pure Lugol's solution for soft tissue contrast, for 24 hours in 40 °C as previously performed [11].

The stained samples were placed in Eppendorf tubes and fixed on the turntable with hot glue, and then were scanned using Bruker SkyScan 1172 micro-CT system. The X-ray energy was set to 45 keV with rotation step of 0.4°, pixel size was set to 7.9 μm; each projection image was averaged out of ten frames. Image reconstruction was done with NRecon v. 1.7.3.1 software by Bruker. The dynamic range has been precisely tuned to achieve the best image contrast.

2.5. Positronium lifetime analysis.

The positronium annihilation lifetime spectroscopy (PALS) spectrum, obtained from the distribution of these measured lifetimes, is typically deconvoluted into three distinct exponential lifetime components[63]:

- the shortest lifetime component, approximately 0.125 ns, corresponds to para-positronium (p-Ps) annihilation;
- an intermediate lifetime component, typically ranging from 0.4 to 0.5 ns, is attributed to free positron annihilation;
- the longest lifetime component, denoted as $\tau_3$, which usually ranges from 1 to 4 ns in biological materials, is due to the annihilation of ortho-positronium (o-Ps). [64, 65]

The o-Ps lifetime is extremely sensitive to its surrounding environment, making it a powerful probe. Its measured value can be directly correlated to the mean radius of the free volume (size of the void) within the material, which can be calculated using semi-empirical models, such as the widely recognized Tao-Eldrup model.[66]

In the present study, ~~the~~ coincidences were recorded in a 50 ns time window, and the data were analyzed for ~~the~~ positronium imaging using the criteria as detailed in [41]. The obtained data on the positions and ~~the~~ time differences between ~~the~~ annihilation and prompt photon emission, as described in [41] are presented in the histograms shown in section Results. The mean ortho positronium lifetime was obtained by fitting a superposition of Gaussian convolutions and exponential functions in Origin Lab software with Equation 1 for only data in positive range.

$$y = y_0 + f(x) * h(x) = y_0 + \sum_{i=1}^{4} \frac{A_i}{2t_i} e^{-\left(\frac{x}{t_i}\right)^2 - \frac{x-x_c}{t_i}} \int_{-\infty}^{z_i} \frac{1}{\sqrt{2\pi}} e^{-\frac{\tau^2}{2}} d\tau \qquad (1)$$

where: $f(x) = \sum_{i=1}^{4} \frac{A_i}{2t_i} e^{-\frac{x}{t_i}}$, $h(x) = \frac{1}{2\sqrt{2\pi}w} e^{-\frac{(x-x_c)^2}{w^2}}$ and $z_i = \frac{x-x_c}{w} - \frac{w}{t_i}$

*Where: $y_0$ - background, $A_i$ – normalization coefficient, w – sigma of Gauss function, $x_c$ – offset position of resolution function, $t_i$- positronium lifetime of i component.*

Our experiment considered three main components: short lifetime accounting for p-Ps, lifetime of direct annihilations and o-Ps lifetime of the sample. Additionally, for the fourth component, the lifetime was identified for the source vial of the empty 5 ml Eppendorf presented in Figure 3, which was used with fixed lifetime value in the analysis of the phantom data. CRM material was considered as quality control for measurements. After correctly fitting superposition from Equation 1 with the known positronium lifetime of 1.62(5) ns we could identify time resolution from system with the $^{44}Sc$ source (Figure 3A).

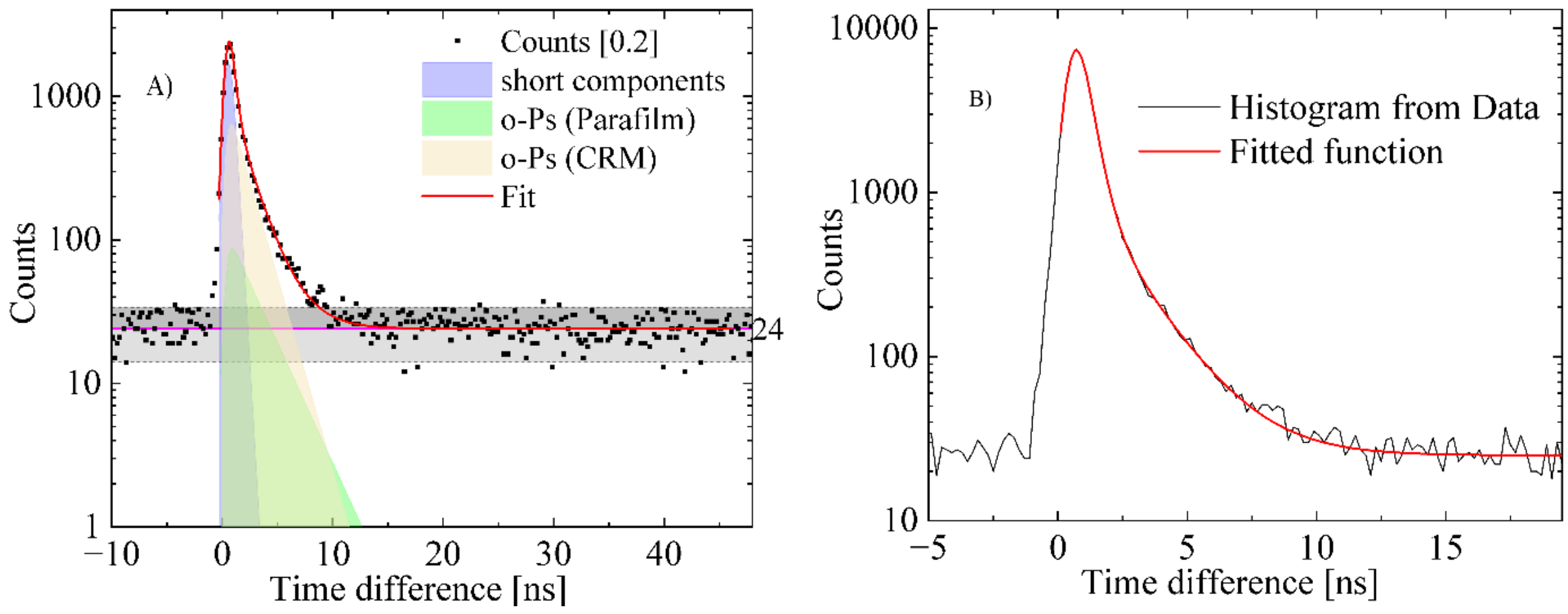

Figure 3. Representation of the main components fitted to experimental data from $^{44}Sc$ phantom measurements. (A) Time difference graph for CRM (fused silica) with all fitted components and 2 SD range of background in grey. The meaning of colors and lines is described in the legend. (B). Time difference histogram for a 0.2 ns bin width measured for a plastic PCR vial with $^{44}Sc$ source inside a 5 ml vial.

Experiment presented in Figure 3B was performed separately to identify o-Ps component from PCR tube in experimental setup.

# 3. Results

## 3.1. Analysis of data for CRM and empty vial samples

Analysis of the data for the CRM and the empty vial revealed high signal intensity from the fused silica at 42% and 8% from the parafilm, with o-Ps lifetime of 2.52 ± 0.08 ns. The identified time resolution is 338(1) ps. The next parameter that needs to be checked is related to the vial that contains the source. Therefore, the source vial (a PCR tube) was placed in an empty 5 ml Eppendorf tube to measure the lifetime, which was later fitted as a component from the source results shown in Figure 3B.

The results of fitting the superposition of the two-component convolution of Gaussian and exponential functions are presented in Table. 1.

Table 1. Results for the scandium source vial.

| **Events** | **Sigma [ns]** | **Short components Intensity [%]** | **Short components lifetime [ns]** | **o-Ps Intensity [%]** | **o-Ps Lifetime [ns]** |
|---|---|---|---|---|---|
| 50992 | 0.348(2) | 77(1) | 0.393(9) | 23(1) | 1.95(10) |

3.2 Ex-vivo experiment with formaldehyde-fixed human tissues and XAD4 material

The annihilation point density distribution images are presented in Figure 4. The locations of human adipose tissue (X= -9.85 cm, Y= -11.05 cm), human cardiac myxoma tissue (X= -9.85 cm, Y= 9.35 cm), human thrombi blood clot (X = 9.95 cm, Y = 9.65 cm), and Amberlite XAD4 (X= 9.95 cm, Y = -10.45 cm) are clearly visible. The weaker signal from certified CRM material (X = -10 cm, Y =13.5 cm) is due to the fact that it was placed in the scanner only for 180 minutes and it was far from the center of the scanner, where the efficiency is lower compared to the imaging efficiency of other samples that were closer to the center. The images projections (Figure 5) clearly show the signal from the CRM material positioned at X= -10 cm, Y= 13.6 cm, Z =14.9 cm.

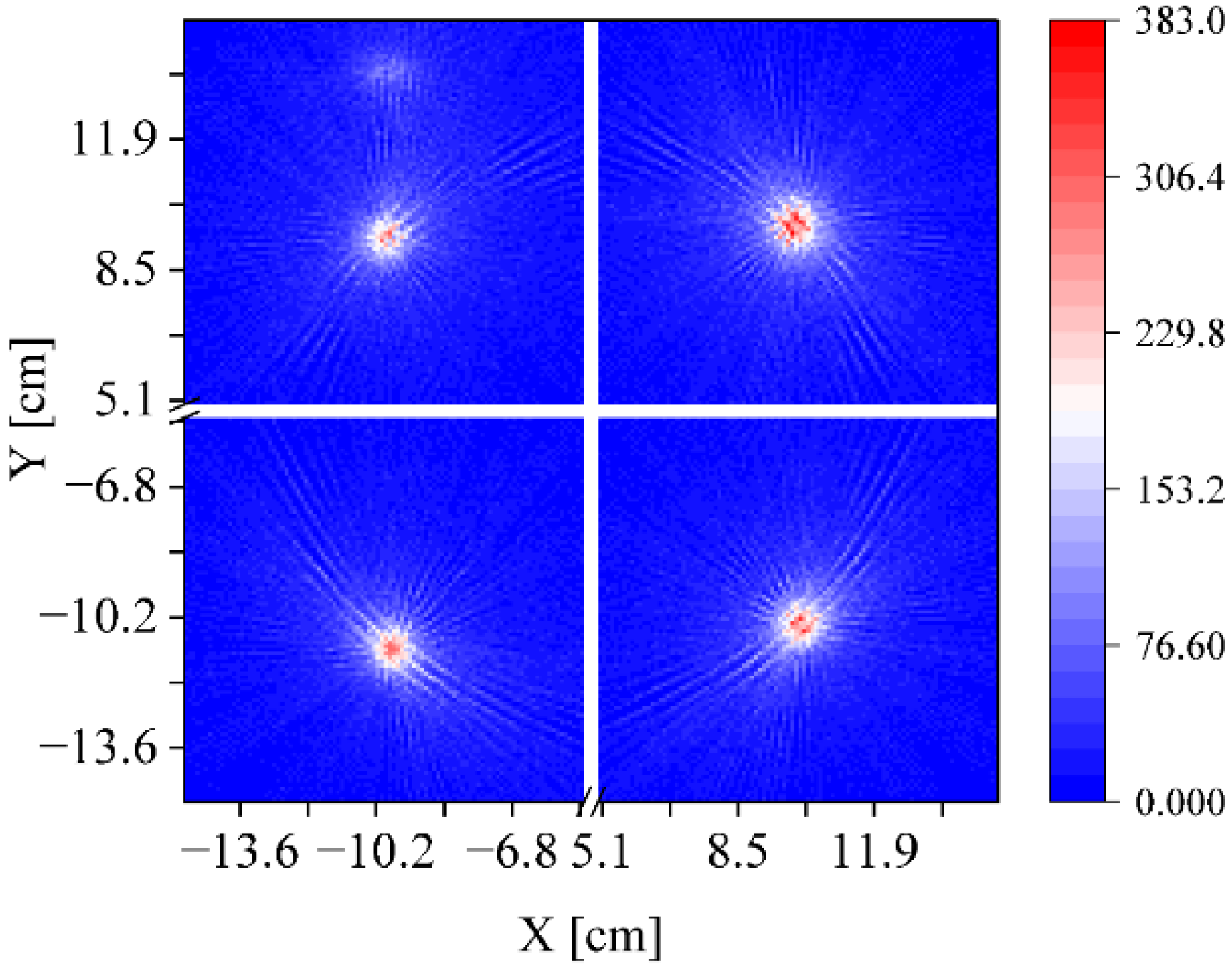

Figure 4. XY plane of the annihilation point image of [0.1 cm bin width].

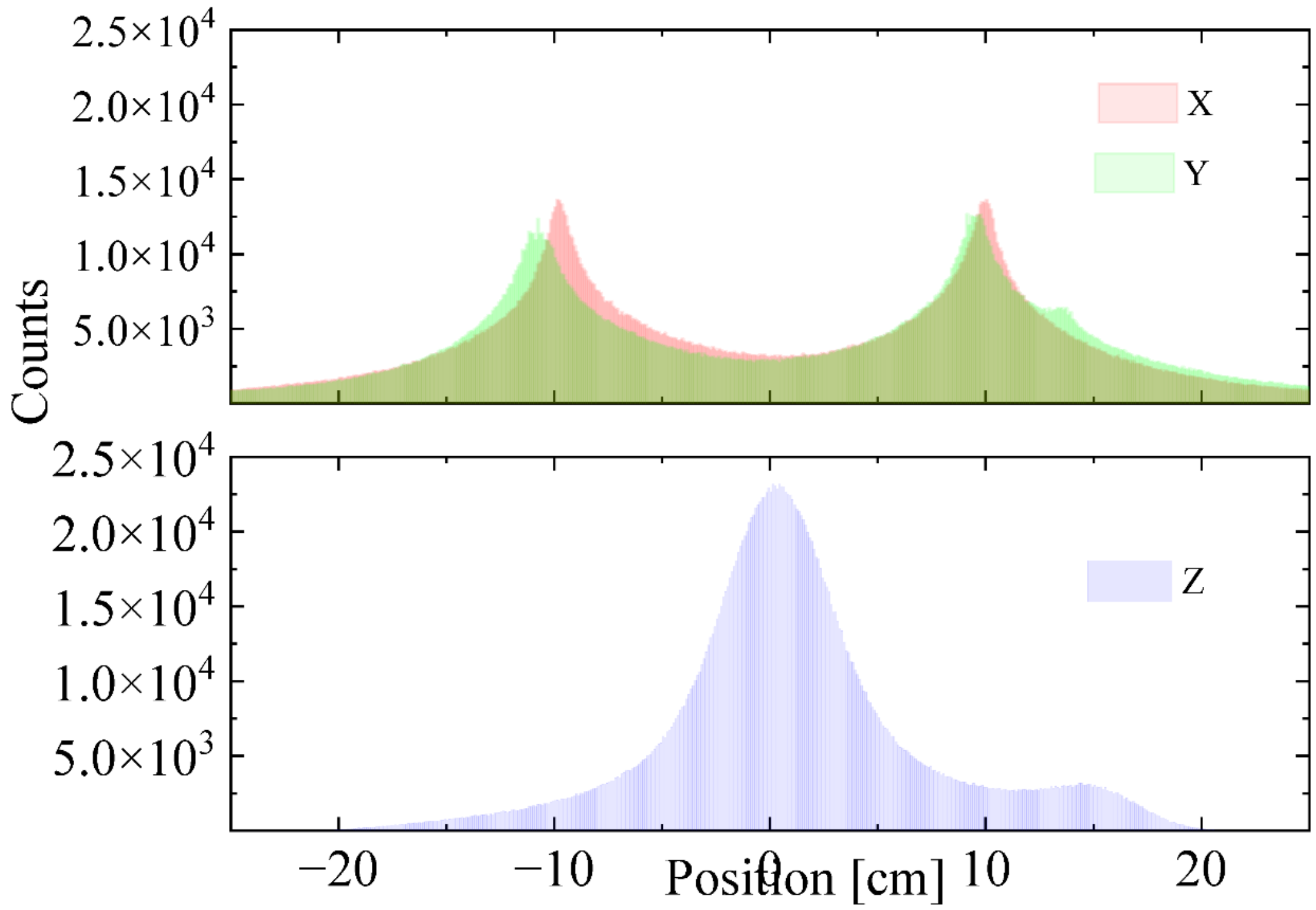

Figure 5. Projections of the annihilation point image. (X) major picks for X= -9.8 cm represent adipose tissue and cardiac myxoma tissue positions and 10 cm represent XAD4 and thrombus positions, (Y) Y= -10.8 cm represent adipose tissue and XAD4 positions and 9.7 cm represent cardiac myxoma tissue and thrombus positions with minor pick on 13.5 cm represent CRM position, (Z) Z = 0.3 cm represent phantoms position and minor at 14.5 cm represent CRM position with 0.1 cm bin width of histogram.

Analysis of the selected data was first performed for adipose and cardiac myxomas, where the source was not in direct contact with sample surface.

The geometric regions used for signal selection from the samples (selected based on the histograms shown in Figure 5) are presented in Table 2. The lifetime spectra from the region of interest were determined with a bin width of 0.2 ns. Equation 1 was fitted to the lifetime spectra.

Table 2. Geometric criteria for selecting events used to determine lifetime spectra.

| | **X plane cut [cm]** | **Y plane cut [cm]** | **Z plane cut [cm]** | **Real position [cm]** |
|---|---|---|---|---|
| Adipose tissue | -11.8: -7.8 | -13: -9 | -5:5 | -10; -10;0 |
| Cardiac myxoma tissue | -12: -8 | 7.4:11.4 | -5:5 | -10;10;0 |
| Thrombi blood clot | 8:12 | 7.7:11.7 | -5:5 | 10;10;0 |
| XAD4 | 8:12 | -12.3: -8.3 | -5:5 | 10; -10;0 |
| CRM (fused silica) | -12.5: -7.5 | 11.6:15.6 | 10:23 | -10;14;15 |

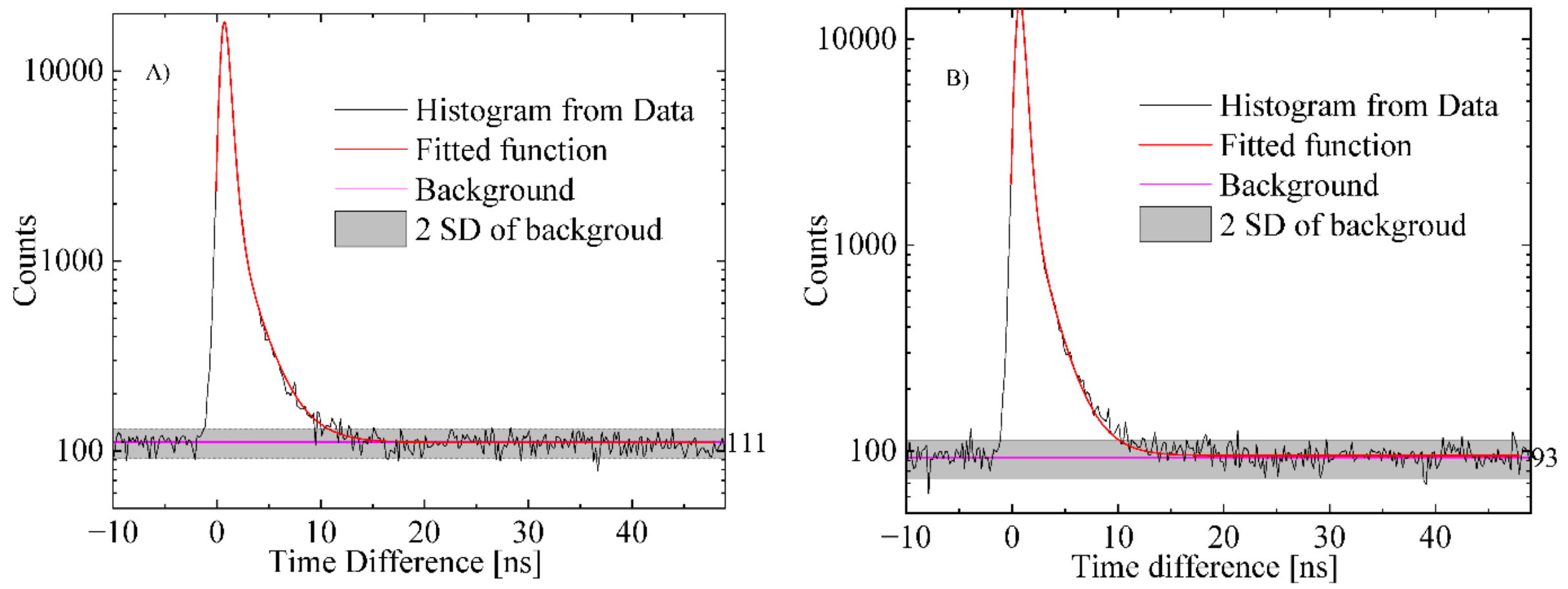

Figure 6. Results of the positronium lifetime analysis for the phantoms on the left. (A) for adipose tissue and (B) cardiac myxoma tissues with a bin width of 0.2 ns. The meaning of the lines is explained in the legend. SD denotes standard deviation.

Figures 6 and 7 present the lifetime spectra determined for adipose tissue, cardiac myxoma, blood clots and the porous polymer XAD4. The result of fitting Equation 1 to these spectra is given in Table 3. As expected, we can observe significant differences in the mean positronium lifetimes between samples. The obtained values of 2.82(22) ns, 1.96(5) ns, 1.82(9) ns for adipose tissue, cardiac myxoma and thrombi, respectively, are consistent with the values of 2.72 ns, 1.92 ns and 1.90 ns determined in previous experiments [11, 67]. Since, cardiac myxoma has the same lifetime as the vial there was no fourth component and the value for p-Ps was screened by the high intensity of the second component showing intensity value lower than its error.

Table 3. Results of analysis of lifetimes for phantoms.

| **Sample** | **Sigma [ns]** | **First Int. [%]** | **First lifetime [ns]** | **Second Int. [%]** | **Second lifetime [ns]** | **Third Int. [%]** | **Third Lifetime [ns]** | **Fourth Int. [%]** |
|---|---|---|---|---|---|---|---|---|
| Adipose tissue | 0.344(2) | 8(2) | 0.052(2) | 71(2) | 0.416(4) | 8.2(1) | 2.82(22) | 12.5(1) |
| Cardiac myxoma tissue | 0.345(1) | - | - | 77.1(1) | 0.389(5) | 22.9(1) | 1.96(5) | - |
| Blood clot | 0.356(2) | 23(2) | 0.055(1) | 55(2) | 0.424(9) | 11.9(7) | 1.82(9) | 10.4(2) |
| XAD 4 | 0.361(2) | 24(2) | 0.060(1) | 55(2) | 0.435(7) | 5.7(4) | 4.49(29) | 14.7(5) |

As shown in Table 3, the sigma value difference is likely due to the increase in scattering potential in phantoms due to the larger number of elements with which positronium could have interacted compared to CRM sample.

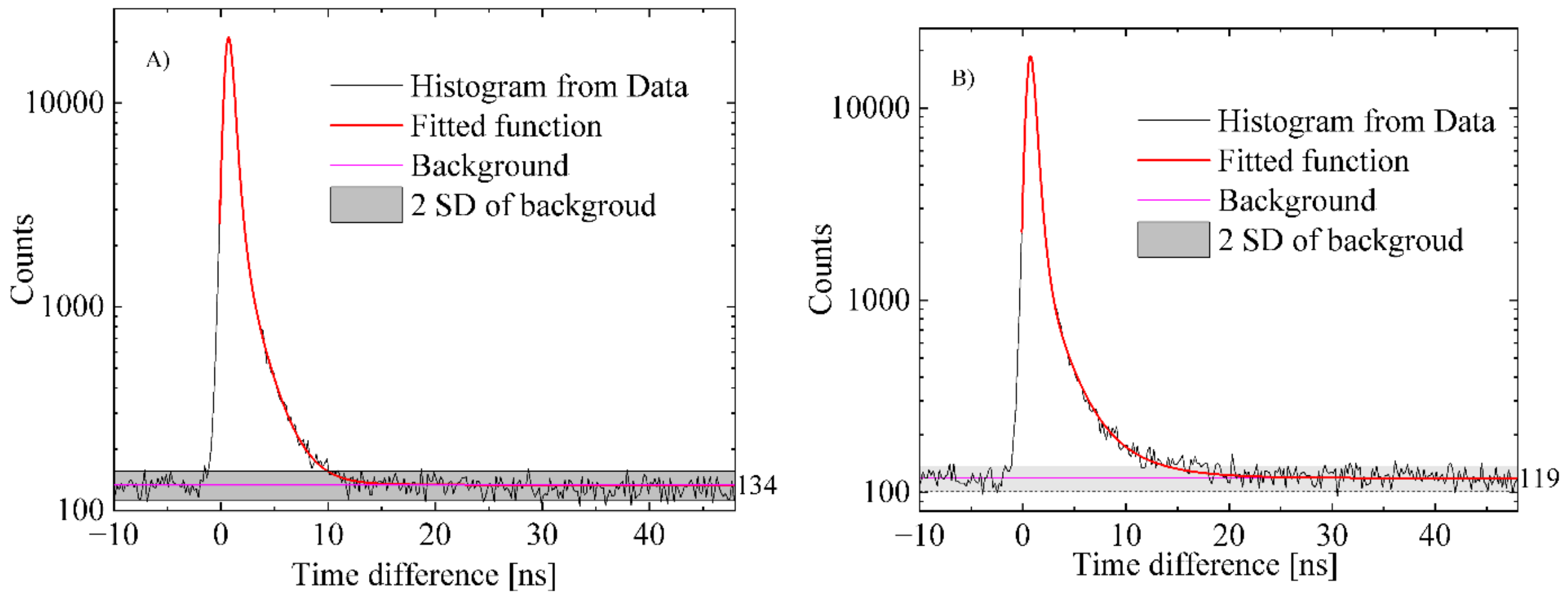

Figure 7. Results of the positronium lifetime analysis for the phantoms on the right. (A) for blood clot and (B) XAD4 with a bin width of 0.2 ns. The meaning of lines is explained in the legend. SD denotes standard deviation.

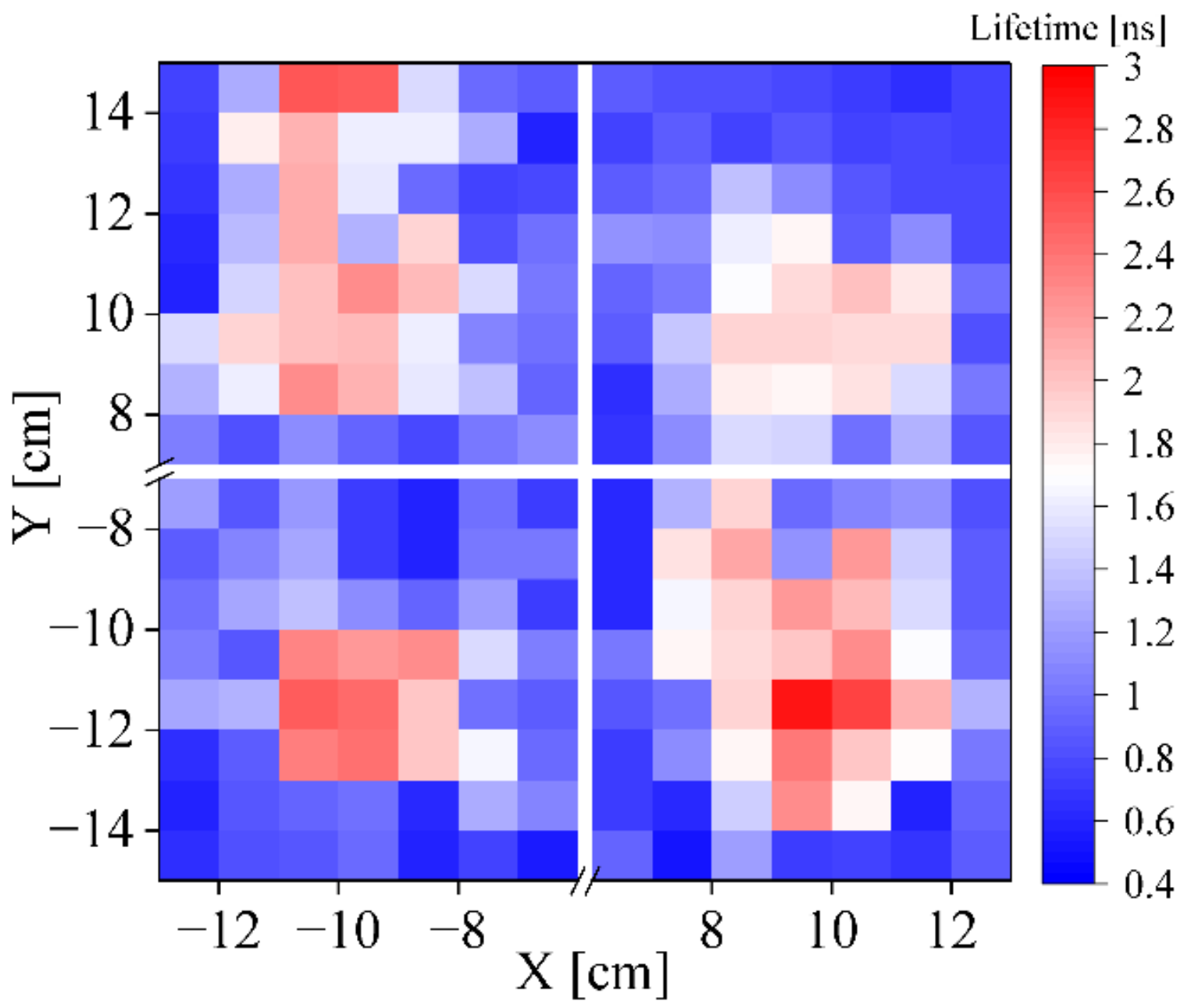

Figure 8. XY plane heatmap of o-Ps lifetimes for biological phantoms and CRM material with one-by-one cm bins.

Figure 8 shows the positronium lifetime image reconstructed for a pixel size of 1 cm x 1 cm. The lifetime spectrum was determined for each voxel a and fitted with Equation 1.

After the radioisotope experiment was completed and the samples did not show external radioactivity, micro-CT scanning was performed to access placement of PCR tubes in phantoms shown in Figure 9 with the same order as in Figure 2.

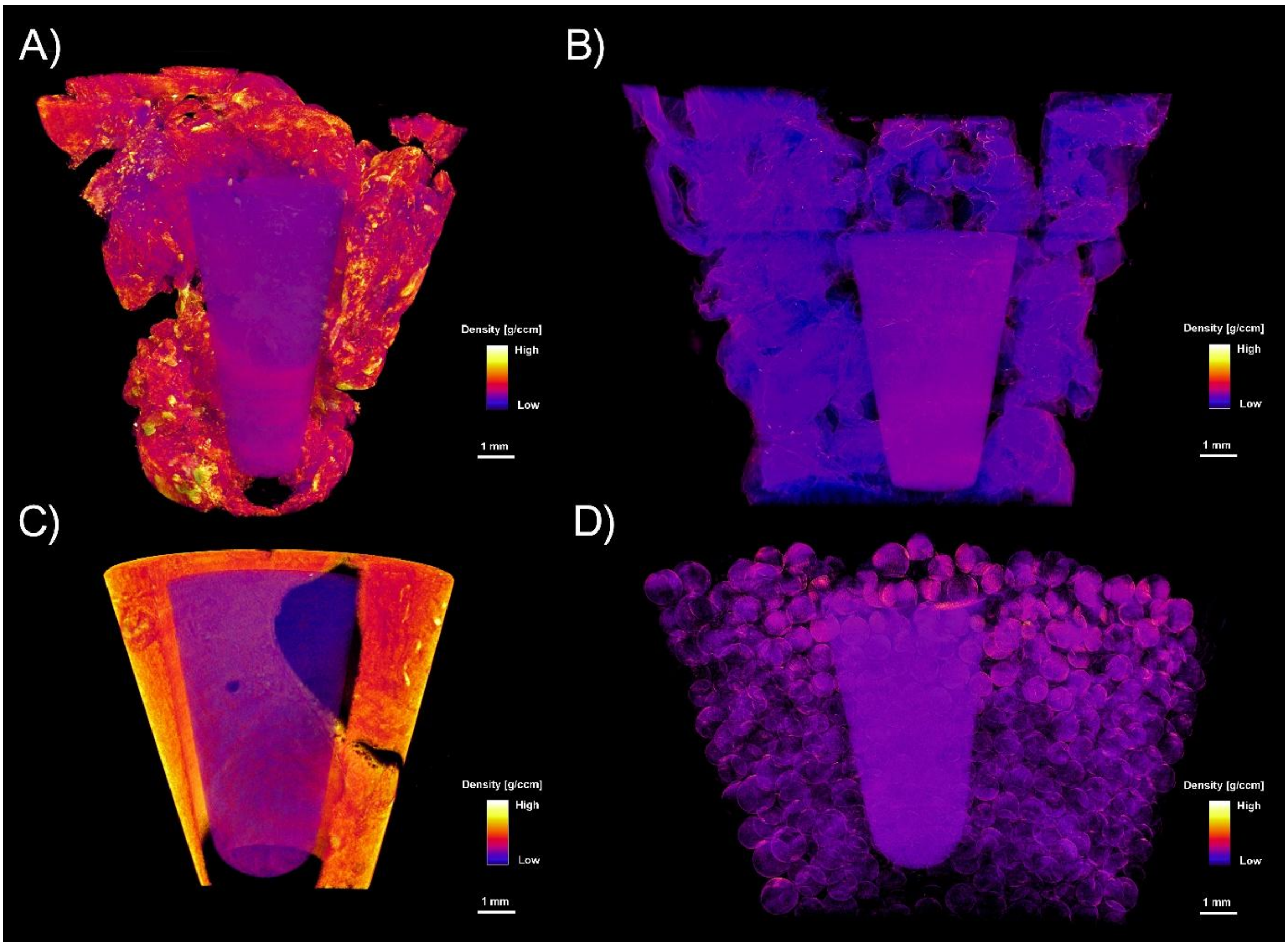

Figure 9. Micro-CT images of the phantoms. In this reconstruction, the scale is uniform throughout. Density [g/ccm] indicates physical density in grams per cubic centimeter - the lighter the color, the denser the tissue/material. (A) ~~this~~ is ~~a~~ myxoma - the most heterogeneous, containing fibrosis/initial mineralization - in yellow and white. (B) adipose tissue, (C) thrombus, and (D) XAD-4. The volume of the imaged biological sample is clearly visible: A: 179.8 mm$^3$; B: 357.8 mm$^3$; C: 80.1 mm$^3$ were chosen arbitrarily.

Heatmapping the positronium lifetime results for individual samples allowed us to determine the homogeneity of these samples (Figure 8). The most uniform distribution is visible for adipose tissue in lower left quadrant, and values with high lifetimes exceeding the physical sample position, particularly in the Y plane from 4 to 8 cm compared to other samples, were observed for XAD4 sample in the lower right quadrant. Compared to the reconstructed micro-CT images (Figure 9), all source solutions were completely immersed in the samples, providing the shortest path for positrons and positronium to interact with samples. It is important to emphasize that all determined lifetimes were correctly identifiable; even the parafilm component approx. 2.6 ns (red) in upper part of higher right quadrant contrasted well with the position of the CRM material with lifetime 1.62 ns (light blue).

## 4. Discussion

Our experiment of first ex-vivo imaging of human tissue using scandium confirmed that the $^{44}$Sc isotope is an ideal candidate as a PLI source characterized by high statistics and low positronium imaging background. In our first in-vivo positronium imaging, 8 MBq of $^{68}$Ga[13] was used, the data were collected for 13 min with a rate of ~150 identified events/min. In the experiment presented in this manuscript we used 7.08 MBq of $^{44}$Sc radionuclide and data were collected for 360 min, which resulted in a high statistic of 3.96 mln events, corresponding to the mean rate of 11000 useful events/min. Thus, as expected, a 70-fold increase in rate was observed in favor of $^{44}$Sc. The observed increase in count rate was similar to one presented in PennPET Explorer experiment [29].

In our experiment, the o-Ps lifetime components were clearly identified with an intensity of approximately 20% [Table 3]. Also in the experiment, we were able to correctly confirm the difference in o-Ps lifetime between adipose and cardiac myxoma tissues [11] and for thrombi [67]. In the case of the XAD-4 sample, only the o-Ps component of 4.5 ns was confirmed [68, 69].

The presented results were obtained for the prototype J-PET scanner with a single layer of modules, 50 cm long [13]. For a total-body J-PET system and with a length of 250 cm, the sensitivity should increase to 28 -fold[13]. It should be emphasized that currently, in addition to the J-PET scanner, 5 other PET systems are capable of positronium imaging: the Prism-PET scanner in New York City, USA [28], the Biograph Vision Quadra in Bern, Switzerland [20, 22, 70], the PennPET Explorer in Philadelphia, USA [29, 71–73], the NeuroEXPLORER (NX) brain PET scanner in USA [74] and the brain-dedicated TOF-PET scanner VRAIN in Chiba, Japan [75, 76].

## 5. Conclusion

This study presents the first high statistic ex-vivo positronium imaging of human tissues using the modular J-PET scanner with the $^{44}$Sc radionuclide.

The potential of the radionuclide $^{44}$Sc for positronium lifetime imaging (PLI) has been clearly demonstrated, especially when comparing the possible increase in data volume compared with the data obtained with the radionuclide $^{68}$Ga, even for small samples with a volume of 2-3 cubic centimeters.

The isotope $^{44}$Sc was used for positronium imaging of a phantom composed of human adipose tissue, cardiac myxoma tissue, a thrombi, as well as a porous polymer XAD4, and a certified reference material (CRM) made from fused silica.

The effectiveness of J-PET for positronium imaging with $^{44}$Sc was validated by reconstruction of the mean ortho-positronium lifetime for CRM material and XAD-4 polymer. The mean ortho-positronium (o-Ps) lifetimes determined for adipose tissue, cardiac myxoma tissues and thrombi were consistent with the results of previous experiments.

**Acknowledgements**

We acknowledge support from the National Science Centre of Poland through grants MAESTRO no. 2021/42/A/ST2/00423 (P.M.), OPUS no. 2021/43/B/ST2/02150 (P.M.),

OPUS24+LAP no. 2022/47/I/NZ7/03112 (E.S.), SONATA no. 2023/50/E/ST2/00574 (S.S.), the Ministry of Science and Higher Education through grant no. IAL/SP/596235/2023 (P.M.), the SciMat, qLife Priority Research Areas budget under the program Excellence Initiative – Research University at Jagiellonian University (P.M. and E.S.) and European Union within the Horizon Europe Framework Programme (ERC Advanced Grant POSITRONIUM no. 101199807), the Research Support Module as part of the Excellence Initiative – Research University program at Jagiellonian University (M.D.). We also acknowledge Polish high-performance computing infrastructure PLGrid (HPC Center: ACK Cyfronet AGH) for providing computer facilities and support within computational grant no. PLG/2024/017688.

The authors declare the following commercial interests/personal relationships which may be considered as potential competing interests with the work reported in this paper: Paweł Moskal is an inventor on a patent related to this work. [Patent nos.: (Poland) PL 227658, (Europe) EP 3039453, and (United States) US 9,851,456], filed (Poland) 30 August 2013, (Europe) 29 August 2014, and (United States) 29 August 2014; published (Poland) 23 January 2018, (Europe) 29 April 2020, and (United States) 26 December 2017. The authors would like to thank Drs J. Stępniewski and G. Grudzień for providing clinical material for the study.

Other authors declare that they have no known conflicts of interest in terms of competing commercial interests or personal relationships that could have an influence or are relevant to the work reported in this paper.